\documentclass[11pt]{article}
\usepackage{amsmath,amssymb,amsthm,amsxtra,overpic,bbm,bm,epsfig,ulem,color,multirow}
\usepackage{braket}

\begin{document}

\begin{center}
{\Large Purely flavon driven leptogenesis \\for exactly degenerate Dirac or Majorana neutrino mass matrix \\ in the type-I seesaw model}
\end{center}

\vspace{0.05cm}

\begin{center}
{\bf Yan Shao, \bf Zhen-hua Zhao\footnote{Corresponding author: zhaozhenhua@lnnu.edu.cn}} \\
{ $^1$ School of Physics and Electronic Technology, Liaoning Normal University, Dalian 116029, China \\
$^2$ Center for Theoretical and Experimental High Energy Physics, \\ Liaoning Normal University, Dalian 116029, China }
\end{center}

\vspace{0.2cm}

\begin{abstract}

While the type-I seesaw model provides an attractive framework for simultaneously explaining the origin of neutrino masses and the baryon asymmetry of the Universe, flavor symmetries offer a natural approach to understanding the observed neutrino mixing pattern. However, the highly constrained neutrino mass structures predicted by flavor symmetries may also prevent conventional leptogenesis to work. In particular, when the Dirac neutrino mass matrix $M^{}_{\rm D}$ or the Majorana right-handed-neutrino (RHN) mass matrix $M^{}_{\rm R}$ is exactly degenerate (namely, they are proportional to the identity matrix $I$), the CP asymmetry of RHN decays will vanish either due to the orthogonality of different neutrino Yukawa coupling columns or the exact degeneracy of RHN masses. In this work, we propose that the flavon fields (which are inherently present in flavor-symmetry neutrino mass models to be responsible for generating the nontrivial neutrino flavor structure) can naturally overcome these obstacles. For the case of $M^{}_{\rm D} \propto I$, flavon--RHN interactions induce additional decay channels $N_I \to N_J \phi$ and generate new CP-violating contributions to RHN decays. For the case of $M^{}_{\rm R} \propto I$, three-body decays $N_I\rightarrow L^{}_\alpha H\phi$ mediated by heavy vectorlike fermions provide a CP asymmetry source even when the RHN masses are exactly degenerate. Our results demonstrate that the flavon sector provides a natural connection between neutrino flavor structure and leptogenesis, offering a new mechanism for leptogenesis in flavor-symmetry-based neutrino mass models.

\end{abstract}

\newpage

\section{Introduction}

The Standard Model (SM) of particle physics has achieved remarkable success in describing most experimental observations. Nevertheless, it fails to account for two fundamental phenomena: the origin of neutrino masses and the observed baryon asymmetry of the Universe (BAU). On the one hand, neutrino oscillation experiments have firmly established that neutrinos are massive and mixed states~\cite{xing}, in clear contradiction with the SM prediction of massless neutrinos. On the other hand, cosmological observations indicate a tiny but non-vanishing excess of baryons over antibaryons, quantified by~\cite{planck}
\begin{eqnarray}
Y^{}_{\rm B}\equiv \frac{n^{}_{\rm B}-n^{}_{\overline{\rm B}}}{s}
\simeq (8.69\pm0.04)\times10^{-11},
\label{1.1}
\end{eqnarray}
where $n^{}_{\rm B}$ ($n^{}_{\overline{\rm B}}$) denotes the baryon (antibaryon) number density and $s$ is the entropy density.

A particularly appealing framework capable of addressing both problems simultaneously is the type-I seesaw model~\cite{seesaw1}-\cite{seesaw5}, in which the SM is extended by three heavy right-handed neutrinos (RHNs) $N_I$ ($I=1,2,3$). The Yukawa interactions connecting the left- and right-handed neutrinos generate the Dirac neutrino mass matrix $M^{}_{\rm D}=Y^{}_\nu v$, where $Y_\nu$ is the neutrino Yukawa coupling matrix and $v=174$ GeV is the Higgs vacuum expectation value (VEV), while the RHNs possess a Majorana mass matrix $M^{}_{\rm R}$. Under the seesaw condition $M^{}_{\rm R}\gg M^{}_{\rm D}$, the effective light-neutrino mass matrix is given by
\begin{eqnarray}
M^{}_\nu=-M^{}_{\rm D}M^{-1}_{\rm R}M^{T}_{\rm D}.
\label{1.2}
\end{eqnarray}
This framework naturally explains the smallness of neutrino masses and simultaneously provides an elegant explanation for the origin of the BAU through leptogenesis~\cite{leptogenesis}-\cite{Lreview4}, where the CP-violating decay of RHNs generates a lepton asymmetry that is partially converted into baryon asymmetry through sphaleron processes.

The observed particular neutrino mixing pattern suggests the possible existence of an underlying flavor structure in the lepton sector, and non-Abelian discrete flavor symmetries, such as the $A_4$ and $S_4$ groups, provide an attractive framework for understanding its origin~\cite{FS1}-\cite{FS4}. In many type-I seesaw models based on such symmetries, the three lepton doublets and the three RHNs are assigned to triplet representations. In the exact flavor-symmetric limit, this assignment can lead to degenerate neutrino mass matrices: either the Dirac neutrino mass matrix or the RHN mass matrix may take the flavor-symmetric form~\cite{FS1}-\cite{FS4}
\begin{eqnarray}
M^{}_{\rm D}=m^{}_{\rm D}
\left(
\begin{array}{ccc}
1&0&0\\
0&0&1\\
0&1&0
\end{array}
\right)
\qquad {\rm or}\qquad
M^{}_{\rm R}=M_0
\left(
\begin{array}{ccc}
1&0&0\\
0&0&1\\
0&1&0
\end{array}
\right),
\label{1.3}
\end{eqnarray}
where $m^{}_{\rm D}$ and $M_0$ denote their overall mass scales. Since the exact flavor-symmetry limit generally cannot reproduce the observed neutrino mixing pattern, the flavor symmetry must be broken appropriately. A simple and natural possibility is that one of $M^{}_{\rm D}$ and $M^{}_{\rm R}$ remains in the exactly degenerate form shown in Eq.~(\ref{1.3}), while the other acquires a nontrivial flavor structure from the breaking of the flavor symmetry. In the literature, the most common approach to realize this is to spontaneously break the flavor symmetry through specific VEV alignments of the flavon fields~\cite{FS1, King:2011zj}-\cite{Ding:2013hpa}.

Despite their simplicity, both exactly degenerate structures in Eq.~(\ref{1.3}) prevent the conventional leptogenesis mechanism from working. For the exactly degenerate form of $M^{}_{\rm D}$, after transforming to the RHN mass basis, the flavor symmetry enforces orthogonality among different columns of the neutrino Yukawa coupling matrix (see Eq.~(\ref{2.1.5})), causing the CP asymmetries of RHN decays to vanish. For the exactly degenerate form of $M^{}_{\rm R}$, the three RHNs are exactly degenerate in mass, and the CP asymmetries of conventional RHN decays vanish. It is therefore worthwhile to explore whether leptogenesis can be realized while preserving the original exactly degenerate structures.

In this paper, we propose that the flavon fields (which are inherently present in flavor-symmetry neutrino mass models to be responsible for the spontaneous breaking of the flavor symmetry) can naturally help us overcome these problems. In the two exactly degenerate limits considered in this work, either $M^{}_{\rm D}$ or $M^{}_{\rm R}$ retains the form given in Eq.~(\ref{1.3}), while the nontrivial structure of the other mass matrix is generated by flavon fields after spontaneous breaking of the flavor symmetry. Beyond their conventional role in generating neutrino flavor structures, these flavon fields can also participate in leptogenesis as decay products of RHNs, providing the CP-violating sources required for successful leptogenesis. Therefore, leptogenesis naturally emerges from the same flavon sector responsible for generating the neutrino mass structure. This establishes a direct connection between the origin of neutrino flavor structure and the dynamics responsible for the baryon asymmetry.

To demonstrate this mechanism, we consider an $S_4$ flavor framework based on the trimaximal (TM) mixing patterns~\cite{TB2-TM}--\cite{TM-5}, which is consistent with current neutrino oscillation data. In this framework, the flavon fields play a dual role by connecting the realization of the TM1 mixing structure with the leptogenesis mechanism considered in this work. For the case of $M^{}_{\rm D} \propto I$, the flavon--RHN interactions open additional RHN decay channels $N_I \to N_J \phi$ (where $\phi$ denotes a flavon field) and generate new CP-violating contributions to RHN decays. For the case of $M^{}_{\rm R} \propto I$, heavy vectorlike fermions mediate the three-body RHN decays $N_I\rightarrow L^{}_\alpha H\phi$, which provide a non-vanishing CP-asymmetry source even when the RHN masses are exactly degenerate. For both cases, we obtain the generated lepton asymmetries by solving the flavor-dependent Boltzmann equations.

The remaining parts of this paper are organized as follows. In Sections~2 and 3, we separately investigate purely flavon-driven leptogenesis for the exactly degenerate forms of the Dirac and Majorana neutrino mass matrices given in Eq.~(\ref{1.3}), respectively. Finally, a summary of our main results is presented in Section~4.

\section{Leptogenesis for the case of $M^{}_{\rm D} \propto I$}

In this section, we study leptogenesis for the exactly degenerate form of $M^{}_{\rm D}$ given in Eq.~(\ref{1.3}), while the RHN mass matrix $M^{}_{\rm R}$ acquires a nontrivial flavor structure through the spontaneous breaking of the flavor symmetry. For this form of $M^{}_{\rm D}$, the conventional leptogenesis mechanism does not work because the orthogonality among different columns of the neutrino Yukawa coupling matrix (after transforming into the mass basis of RHNs) causes the CP asymmetries of RHN decays to vanish. We therefore investigate how the flavon fields responsible for generating the nontrivial structure of $M^{}_{\rm R}$ can also participate in leptogenesis through their interactions with the RHNs, opening additional RHN decay channels and providing new CP-violating sources. In Section~2.1, we introduce the flavor-symmetry model and derive the new CP-violating contributions to the CP asymmetries of RHN decays. In Section~2.2, we study the corresponding Boltzmann equations including the additional decay processes and present the numerical results for the baryon asymmetry.

\subsection{Model framework and CP asymmetries}

For the exactly degenerate form of $M^{}_{\rm D}$ given in Eq.~(\ref{1.3}), the nontrivial neutrino flavor structure originates from the RHN mass matrix $M^{}_{\rm R}$. The tri-bimaximal (TBM) mixing pattern~\cite{TB1, TB2-TM} provides a simple and predictive framework for neutrino mixing, although its prediction of $\theta^{}_{13}=0$ requires modifications in light of the observed nonzero reactor mixing angle. Among the viable alternatives, the TM1 mixing pattern preserves the first column of the TBM mixing matrix while modifying the remaining columns and provides a simple framework consistent with current neutrino-oscillation data~\cite{global1,global2}. Moreover, the TM1 mixing pattern has attracted renewed phenomenological interest due to precision measurements by JUNO~\cite{JUNO:2025gmd} (see recent studies~\cite{Xing:2025bdm}--\cite{Ardakanian:2026rwz}). Motivated by these facts, we consider a flavor-symmetry neutrino mass framework in which the RHN mass matrix $M^{}_{\rm R}$ acquires a nontrivial flavor structure that leads to the TM1 mixing pattern.

To realize the TM1 mixing pattern, we consider an $S_4$ flavor-symmetry neutrino mass framework following the flavon vacuum alignments given in Ref.~\cite{Luhn:2013lkn}. The three RHNs $N_I$ ($I=1,2,3$) are assigned to the triplet representation ${\bf3}$ of $S_4$, and four flavon fields $\Phi_1$, $\Phi_2$, $\Phi_3$, and $\Phi_4$ are introduced, transforming as ${\bf1}$, ${\bf2}$, ${\bf3}'$, and ${\bf3}'$, respectively. In this framework, the vacuum alignments of flavons transforming as ${\bf1}$, ${\bf2}$, and ${\bf3}'$ generate the structure associated with TBM mixing in the RHN mass matrix, while an additional ${\bf3}'$ flavon with a different vacuum alignment induces the deviation from the TBM structure required for TM1 mixing. The relevant RHN--flavon interactions are given by
\begin{eqnarray}
-\mathcal{L}_{N\phi}=
\frac{1}{2}
\left[
\alpha_1(NN)_{\bf 1}\Phi_1
+\alpha_2(NN)_{\bf 2}\Phi_2
+\alpha_3(NN)_{{\bf 3}'}\Phi_3
+\alpha_4(NN)_{{\bf 3}'}\Phi_4
\right]
+{\rm h.c.},
\label{2.1.1}
\end{eqnarray}
where $\alpha_k$ ($k=1,2,3,4$) denotes the dimensionless coupling associated with the flavon field $\Phi_k$. For flavon multiplets transforming under non-trivial representations of $S_4$, we denote the components of $\Phi_k$ by $\phi_{k,\rho}$, with $\rho$ labeling the corresponding flavon components. Their coupling structures are determined by the relevant $S_4$ Clebsch--Gordan coefficients~\cite{King:2011zj}. After the spontaneous breaking of $S_4$, the flavon fields acquire the vacuum alignments
\begin{eqnarray}
\langle \Phi_1\rangle=v_1,\qquad
\langle \Phi_2\rangle=v_2(1,1)^T,\qquad
\langle \Phi_3\rangle=v_3(1,1,1)^T,\qquad
\langle \Phi_4\rangle=v_4(0,1,-1)^T ,
\label{2.1.2}
\end{eqnarray}
where $v_k$ denote the corresponding VEV parameters. Substituting these vacuum alignments into Eq.~(\ref{2.1.1}), the Majorana RHN mass matrix takes the form
\begin{eqnarray}
\begin{aligned}
M^{}_{\rm R}={}&
\alpha^{}_1 v^{}_1 \begin{pmatrix} 1&0&0\\ 0&0&1\\ 0&1&0 \end{pmatrix} +\alpha^{}_2 v^{}_2 \begin{pmatrix} 0&1&1\\ 1&1&0\\ 1&0&1\end{pmatrix} +\alpha^{}_3 v^{}_3 \begin{pmatrix} 2&-1&-1\\ -1&2&-1\\ -1&-1&2\end{pmatrix}
+ \alpha^{}_4 v^{}_4 \begin{pmatrix} 0&1&-1\\ 1&2&0\\ -1&0&-2 \end{pmatrix}.
\end{aligned}
\label{2.1.3}
\end{eqnarray}
The first three terms are invariant under the $Z_2^S\times Z_2^U$ Klein symmetry, while the last term breaks this symmetry but preserves the residual $Z_2^{SU}$ symmetry. As a result, the full $M^{}_{\rm R}$ preserves the residual $Z_2^{SU}$ symmetry and leads to the TM1 mixing pattern.

In order to obtain the RHN masses and subsequently facilitate the leptogenesis calculations, we go into the mass basis of RHNs. This is done by transforming the RHN mass matrix in Eq.~(\ref{2.1.3}) into a diagonal form via a unitary rotation $V$:
\begin{eqnarray}
V^T M^{}_{\rm R}V={\rm diag}(M_1,M_2,M_3),
\label{2.1.4}
\end{eqnarray}
where $M^{}_I$ denote the physical masses of the three RHNs. In the following analysis, we focus on the normal ordering of light-neutrino masses, i.e. $m_1<m_2<m_3$. For the exactly degenerate form of $M_{\rm D}$ considered here, the seesaw relation implies an inverse correspondence between the light-neutrino and RHN masses, leading to the RHN mass hierarchy $M_1>M_2>M_3$. Meanwhile, the Dirac neutrino mass matrix in the RHN mass basis becomes
\begin{eqnarray}
M^{\prime}_{\rm D}=M^{}_{\rm D}V,
\label{2.1.5}
\end{eqnarray}
where the corresponding neutrino Yukawa coupling matrix is defined as $Y^{\prime}_\nu=M^{\prime}_{\rm D}/v$. For simplicity, the prime notation will be omitted for $Y_\nu$ in the following analysis. The flavon--RHN couplings are also transformed into the RHN mass basis through the same rotation $V$. Since $V$ diagonalizes the full RHN mass matrix rather than the contribution associated with an individual flavon component, the couplings associated with different flavon components generally become non-diagonal in this basis. We denote these couplings as $\alpha^{(k,\rho)}_{IJ}$, which are determined by the underlying $S_4$ contractions and the rotation to the RHN mass basis, rather than treated as independent parameters. The non-diagonal elements of $\alpha^{(k,\rho)}_{IJ}$ open additional decay channels $N_I\rightarrow N_J\phi_{k,\rho}$ and provide new CP-violating contributions to RHN decays at the loop level ~\cite{LeDall:2014too}. Similar scalar-assisted RHN decay processes and their contributions to leptogenesis have been studied in Refs.~\cite{LeDall:2014too}-\cite{Spalding:2026otj}. In the present framework, these scalar interactions arise from the flavon fields responsible for generating the nontrivial flavor structure of $M_{\rm R}$ (see also Ref.~\cite{Shao:2026xcb}).

To illustrate the role of these additional interactions, we first recall the conventional contribution to the CP asymmetry of RHN decays. For hierarchical RHN masses, the flavor-dependent CP asymmetry for the decay processes $N_I\to L_\alpha H$ is given by
\begin{eqnarray}
&& \varepsilon^{0}_{I\alpha}
=
\frac{1}{8\pi(Y^\dagger_\nu Y_\nu)_{II}}
\sum_{J\neq I}
\left\{
{\rm Im}\left[
(Y^*_\nu)_{\alpha I}(Y_\nu)_{\alpha J}
(Y^\dagger_\nu Y_\nu)_{IJ}
\right]
{\cal F}\left(\frac{M_J^2}{M_I^2}\right)
\right.
\nonumber\\
&&\hspace{1.0cm}
\left.
+
{\rm Im}\left[
(Y^*_\nu)_{\alpha I}(Y_\nu)_{\alpha J}
(Y^\dagger_\nu Y_\nu)^*_{IJ}
\right]
{\cal G}\left(\frac{M_J^2}{M_I^2}\right)
\right\},
\label{2.1.6}
\end{eqnarray}
with $Y^{}_{\nu}=M^{\prime}_{\rm D}/v$, ${\cal F}(x) = \sqrt{x} {(2-x)/(1-x)+ (1+x) \ln [x/(1+x)] }$ and ${\cal G}(x) = 1/(1-x)$. This contribution originates from the interference between the tree-level and one-loop contributions to RHN decays. For the exactly degenerate form of $M_{\rm D}$ in Eq.~(\ref{1.3}), $M^{\prime}_{\rm D}$ in Eq.~(\ref{2.1.5}) is simply proportional to the unitary matrix $V$ that diagonalizes $M^{}_{\rm R}$ (up to a column permutation). This relation leads to the orthogonality among different columns of $Y^{}_{\nu}$, i.e., $(Y^{\dagger}_{\nu}Y^{}_{\nu})^{}_{IJ}=0$, which results in a vanishing $\varepsilon^{0}_{I\alpha}$ in Eq.~(\ref{2.1.6}). Therefore, the conventional leptogenesis mechanism does not work.

\begin{figure*}
\centering
\includegraphics[width=5.5in]{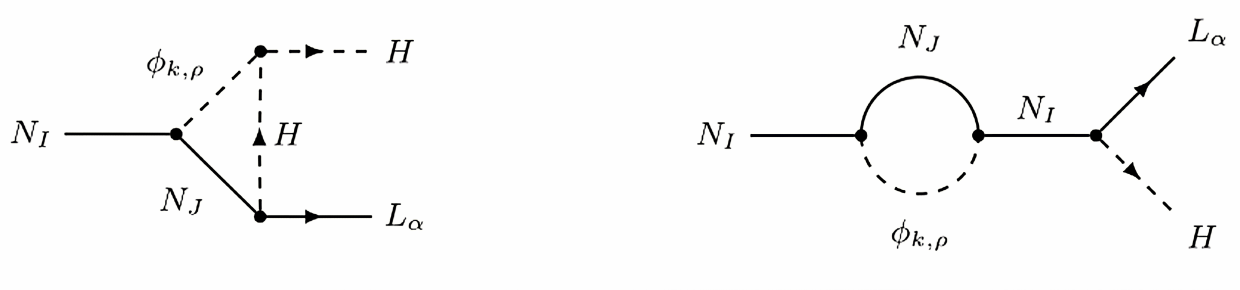}
\caption{Vertex and self-energy corrections involving flavons and contributing to the CP asymmetry of $N_I\to L^{}_\alpha H$.}
\label{fig1}
\end{figure*}

In the model considered here, the nonzero flavon--RHN couplings $\alpha^{(k,\rho)}_{IJ}$ generate additional vertex and self-energy contributions involving $\phi^{}_{k,\rho}$ to the flavor-dependent CP asymmetries of $N_I$ decays (see Figure~\ref{fig1})~\cite{LeDall:2014too}. The total CP asymmetry can therefore be written as
\begin{eqnarray}
\varepsilon_{I\alpha}
=
\varepsilon^{0}_{I\alpha}
+
\sum_{k,\rho}
\left[
\varepsilon^{v,(k,\rho)}_{I\alpha}
+
\varepsilon^{s,(k,\rho)}_{I\alpha}
\right],
\label{2.1.7}
\end{eqnarray}
where $\varepsilon^{v,(k,\rho)}_{I\alpha}$ and $\varepsilon^{s,(k,\rho)}_{I\alpha}$ denote the additional vertex and self-energy contributions involving $\phi^{}_{k,\rho}$, respectively. They can generally be expressed as
\begin{eqnarray}
&&\varepsilon^{v,(k,\rho)}_{I\alpha}
=
\sum_J
\left[
\frac{
{\rm Im}\left\{
(Y^*_\nu)_{\alpha J}(Y_\nu)_{\alpha I}
\beta_k\alpha^{(k,\rho)}_{IJ}
\right\}
}{
8\pi(Y^\dagger_\nu Y_\nu)_{II}M_I
}
{\cal F}^{v,(k)}_{IJ,R}
\right.
\left.
+
\frac{
{\rm Im}\left\{
(Y^*_\nu)_{\alpha J}(Y_\nu)_{\alpha I}
\beta_k\alpha^{(k,\rho)*}_{IJ}
\right\}
}{
8\pi(Y^\dagger_\nu Y_\nu)_{II}M_I
}
{\cal F}^{v,(k)}_{IJ,L}
\right],
\nonumber \\
&&
\varepsilon^{s,(k,\rho)}_{I\alpha}
=
\sum_{J,K}
\Bigg[
\frac{
{\rm Im}\left\{
(Y^*_\nu)_{\alpha K}(Y_\nu)_{\alpha I}
\alpha^{(k,\rho)}_{KJ}
\alpha^{(k,\rho)*}_{IJ}
\right\}
}{
8\pi(Y^\dagger_\nu Y_\nu)_{II}
}
{\cal F}^{s,(k)}_{IJK,RL}
+
\frac{
{\rm Im}\left\{
(Y^*_\nu)_{\alpha K}(Y_\nu)_{\alpha I}
\alpha^{(k,\rho)*}_{KJ}
\alpha^{(k,\rho)*}_{IJ}
\right\}
}{
8\pi(Y^\dagger_\nu Y_\nu)_{II}
}
{\cal F}^{s,(k)}_{IJK,LL}
\nonumber \\
&&\hspace{1.4cm}
+\frac{
{\rm Im}\left\{
(Y^*_\nu)_{\alpha K}(Y_\nu)_{\alpha I}
\alpha^{(k,\rho)}_{KJ}
\alpha^{(k,\rho)}_{IJ}
\right\}
}{
8\pi(Y^\dagger_\nu Y_\nu)_{II}
}
{\cal F}^{s,(k)}_{IJK,RR}
+
\frac{
{\rm Im}\left\{
(Y^*_\nu)_{\alpha K}(Y_\nu)_{\alpha I}
\alpha^{(k,\rho)*}_{KJ}
\alpha^{(k,\rho)}_{IJ}
\right\}
}{
8\pi(Y^\dagger_\nu Y_\nu)_{II}
}
{\cal F}^{s,(k)}_{IJK,LR}
\Bigg],
\label{2.1.8}
\end{eqnarray}
with
\begin{eqnarray}
&&{\cal F}^{v,(k)}_{IJ,L}
=
-\sqrt{\delta^{(k)}_{JI}}
+r_{JI}\ln G^{(k)}_{JI},
\quad{\cal F}^{v,(k)}_{IJ,R}
=-\sqrt{r_{JI}}\sqrt{\delta^{(k)}_{JI}}
+\sqrt{r_{JI}}\ln G^{(k)}_{JI},
\nonumber\\
&&{\cal F}^{s,(k)}_{IJK,LR}
=
\frac{\sqrt{\delta^{(k)}_{JI}}}{2}
\frac{\sqrt{\delta^{(k)}_{JI}+4r_{JI}}}
{1-r_{KI}},
\quad{\cal F}^{s,(k)}_{IJK,RR}
=
\sqrt{\delta^{(k)}_{JI}}
\frac{\sqrt{r_{JI}}\sqrt{r_{KI}}}
{1-r_{KI}},
\nonumber\\
&&{\cal F}^{s,(k)}_{IJK,LL}
=
\sqrt{\delta^{(k)}_{JI}}
\frac{\sqrt{r_{JI}}}
{1-r_{KI}},
\quad{\cal F}^{s,(k)}_{IJK,RL}
=
\frac{\sqrt{\delta^{(k)}_{JI}}}{2}
\frac{\sqrt{r_{KI}}\sqrt{\delta^{(k)}_{JI}+4r_{JI}}}
{1-r_{KI}},
\nonumber\\
&&
G^{(k)}_{JI}
=
\frac{
\sqrt{\delta^{(k)}_{JI}+4r_{JI}\sigma^{(k)}_I+2\sigma^{(k)}_I}
+\sqrt{\delta^{(k)}_{JI}}
}{
\sqrt{\delta^{(k)}_{JI}+4r_{JI}\sigma^{(k)}_I+2\sigma^{(k)}_I}
-\sqrt{\delta^{(k)}_{JI}}
}
\times
\frac{
\sqrt{\delta^{(k)}_{JI}+4r_{JI}\sigma^{(k)}_I}
-\sqrt{\delta^{(k)}_{JI}}
}{
\sqrt{\delta^{(k)}_{JI}+4r_{JI}\sigma^{(k)}_I}
+\sqrt{\delta^{(k)}_{JI}}
},
\nonumber\\
&&r_{IJ}=\frac{M_I^2}{M_J^2},
\qquad
\sigma^{(k)}_I=\frac{m_{\phi_k}^2}{M_I^2},
\qquad
\delta^{(k)}_{IJ}
=
\left(1-r_{IJ}-\sigma^{(k)}_J\right)^2
-4r_{IJ}\sigma^{(k)}_J.
\label{2.1.9}
\end{eqnarray}
We assume that the components within each flavon multiplet have degenerate masses, $m^{}_{\phi_{k,\rho}}=m^{}_{\phi_k}$. In this limit, the loop functions in Eq.~(\ref{2.1.9}) depend only on the common flavon mass $m_{\phi_k}$, while the dependence on the individual flavon components is retained through the corresponding couplings $\alpha^{(k,\rho)}_{IJ}$. Here, $\beta_k$ denotes the effective dimensional trilinear coupling between the flavon fields and the Higgs doublet, whose origin will be discussed around Eq.~(\ref{2.2.6}). These additional contributions arising from the flavon--RHN interactions provide new sources of CP violation through the couplings $\alpha^{(k,\rho)}_{IJ}$, which are not subject to the orthogonality condition responsible for the vanishing of the conventional CP asymmetry. However, in the unflavored treatment adopted in Ref.~\cite{LeDall:2014too}, the flavor-summed CP asymmetry vanishes for $M_{\rm D}\propto I$ considered here. Therefore, a flavor-dependent treatment is essential.

\subsection{Boltzmann equations and phenomenological analysis}

We now turn to the generation and evolution of the lepton asymmetry. In the following analysis, we focus on the parameter region with a sufficiently small lightest neutrino mass, for which the RHN spectrum satisfies $M_1\gg M_2>M_3$. Owing to this mass hierarchy, $N_1$ effectively decouples from the temperature range relevant for the decays of $N_2$ and $N_3$, and its contribution to leptogenesis can therefore be neglected. We thus focus on the coupled evolution of $N_2$ and $N_3$, with the relevant flavon-related decay channel $N_2\rightarrow N_3\phi_{k,\rho}$ being kinematically allowed.

To describe the evolution of the flavor-dependent lepton asymmetries generated by RHN decays, we extend the Boltzmann equations formulated in Ref.~\cite{LeDall:2014too} by including the flavor dependence of the source and washout terms. In addition to the conventional decay and inverse-decay processes $N_I\to L^{}_\alpha H$, we include the additional decay process $N_2\to N_3\phi^{}_{k,\rho}$ and the dominant $\Delta L=2$ scattering processes $N_I N_J\to HH$. The corresponding evolution equations are
\begin{eqnarray}
&& \frac{{\rm d}N_{N_2}}{{\rm d}z}
=
-\left(D_2+D_{23}\right)\Delta_{N_2}
+D_{23}\Delta_{N_3}
-\Delta_{N_2N_3}S_{N_2N_3\to HH}
-\Delta_{N_2N_2}S_{N_2N_2\to HH},
\nonumber\\
&& \frac{{\rm d}N_{N_3}}{{\rm d}z}
=
-\left(D_3+D_{23}\right)\Delta_{N_3}
+D_{23}\Delta_{N_2}
-\Delta_{N_2N_3}S_{N_2N_3\to HH}
-\Delta_{N_3N_3}S_{N_3N_3\to HH},
\nonumber\\
&& \frac{{\rm d}N_{\alpha}}{{\rm d}z}
=
\sum_{I=2,3}\varepsilon_{I\alpha}D_I\Delta_{N_I}
-W_\alpha N_\alpha .
\label{2.2.1}
\end{eqnarray}
Here the temperature-dependent quantities are defined as
\begin{eqnarray}
&&
\Delta_{N_I}(z)
=
\frac{N_{N_I}(z)}{N_{N_I}^{\rm eq}(z)}-1,
\qquad
\Delta_{N_IN_J}
=
\frac{N_{N_I}N_{N_J}}
{N_{N_I}^{\rm eq}N_{N_J}^{\rm eq}}-1,
\qquad
N_{N_I}^{\rm eq}(z)
=
\frac{z_I^2}{2}\mathcal{K}_2(z_I),
\nonumber\\
&&
D_I(z)
=
K_I\frac{z_I^2}{z}
\frac{\mathcal{K}_1(z_I)}{\mathcal{K}_2(z_I)}
N_{N_I}^{\rm eq}(z),
\qquad
D_{23}(z)
=
K_{23}\frac{z_2^2}{z}
\frac{\mathcal{K}_1(z_2)}{\mathcal{K}_2(z_2)}
N_{N_2}^{\rm eq}(z),
\nonumber\\
&&
W_\alpha(z)
=
\sum_{I=2,3}
\frac{1}{4}K_{I\alpha}
\frac{z_I^4}{z}\,
\mathcal{K}_1(z_I),
\label{2.2.2}
\end{eqnarray}
where $\mathcal{K}^{}_{1,2}(z)$ are the modified Bessel functions, $z\equiv M^{}_{\rm min}/T$ (with $M^{}_{\rm min}=M^{}_3$ for the model considered here) is the dimensionless inverse temperature, and $z^{}_I\equiv M^{}_I/T$ corresponds to the $I$-th RHN. The decay parameters $K^{}_I$ are defined as
\begin{eqnarray}
K^{}_{I} \equiv \sum_{\alpha}^{} K^{}_{I\alpha} = \sum_{\alpha}^{} \frac{\Gamma^{}_{I\alpha}}{H(T=M^{}_I)} \;,
\label{2.2.3}
\end{eqnarray}
where the decay width is given by $\Gamma^{}_{I\alpha} = |(Y^{}_{\rm \nu})_{\alpha I} |^{2} M^{}_I/(8\pi)$, and $H(T) =1.66 \sqrt{g_*}\, T^2/M^{}_{\rm Pl}$ is the Hubble rate, with $g^{}_*$ being the number of relativistic degrees of freedom and $M^{}_{\rm Pl} = 1.22\times 10^{19}$ GeV the Planck mass.
In analogy with $K^{}_I$, $K^{}_{23}$ is defined as
\begin{eqnarray}
K^{}_{23}  \equiv \sum_{ k,\rho}^{} \frac{\Gamma(N_2\to N_3 \phi_{k,\rho})}{H(T=M_2)} \;,
\label{2.2.4}
\end{eqnarray}
with
\begin{eqnarray}
\Gamma(N_2\to N_3\phi_{k,\rho})
=
\frac{|\alpha^{({k,\rho})}_{23}|^2M_2}{16\pi}
\left[
(1+r_{32})^2-\sigma_2
\right]
\sqrt{\delta_{32}},
\label{2.2.5}
\end{eqnarray}
The corresponding expressions for the flavon-mediated scattering functions $S_{N_I N_J\to HH}$ are collected in Appendix~A1. By solving the above set of evolution equations, we obtain the final lepton asymmetry $Y^{}_{L}=\frac{\sum_{\alpha}^{} N^{}_{\alpha}}{27\times7.04}$. The resulting baryon asymmetry is then given by $Y^{}_{\rm B}=cY^{}_{\rm L}$, where $c=-28/79$ denotes the sphaleron conversion efficiency relating the lepton asymmetry to the baryon asymmetry.

\begin{figure*}
\centering
\includegraphics[width=4.5in]{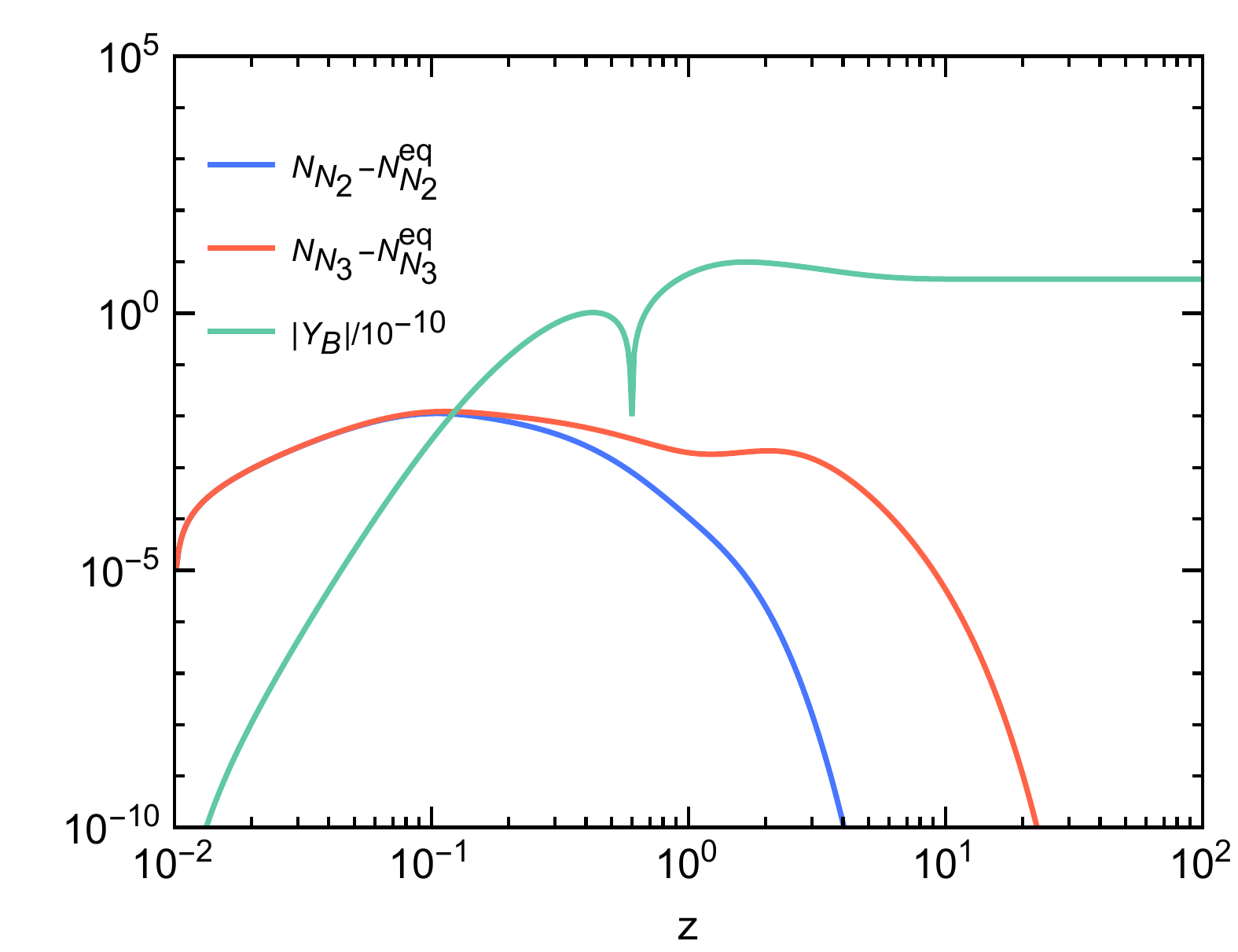}
\caption{ Evolutions of the RHN abundances $N^{}_{N^{}_2, N^{}_3}$ and the baryon asymmetry $Y^{}_{\rm B}$ as functions of $z$ for a representative parameter combination (see the main text for details). }
\label{fig2}
\end{figure*}

Now we are ready to perform the numerical calculations. Owing to the special form of $M^{}_{\rm D}$ in Eq.(\ref{1.3}), taking $m^{}_{\rm D}$ as an input parameter allows us to determine the RHN mass matrix $M^{}_{\rm R}$ in Eq.~(\ref{2.1.3}) from the experimental results for the neutrino parameters. The resulting parameters provide the necessary inputs for solving the Boltzmann equations and computing the baryon asymmetry. For simplicity, we assume that the flavons relevant for leptogenesis share a common mass $m^{}_{\phi_k}=m^{}_\phi$, and that their effective trilinear couplings to the Higgs doublet satisfy $\beta^{}_k=\beta$. As an illustration of the leptogenesis dynamics, we first consider a benchmark parameter choice. We take the characteristic scale of the Dirac neutrino mass matrix to be $m^{}_{\rm D}\sim0.2$~MeV, corresponding to $M_2\simeq4.6$~TeV and $M_3\simeq0.8$~TeV. For this benchmark point, Figure~\ref{fig2} shows the evolutions of the RHN abundances and the resulting baryon asymmetry as functions of $z$ for $|\alpha^{}_i|=0.1$ and $m^{}_\phi=\beta=1$~GeV. At early times ($z\lesssim1$), $N_2$ departs from equilibrium, and its decays, including the $N_2\to N_3\phi^{}_{k,\rho}$ channels, efficiently generate the lepton asymmetry. Around $z\sim1$, the $N_3$-related interactions become significant and partially wash out the previously generated asymmetry. The interplay between asymmetry generation from $N_2$ decays and the subsequent washout processes yields a nonzero baryon asymmetry at late times.

We next examine the theoretical and experimental constraints on the flavon sector in the parameter region that reproduces the observed baryon asymmetry. After the spontaneous breaking of the flavor symmetry, the flavon fields acquire the vacuum alignments given in Eq.~(\ref{2.1.2}) and can mix with the Higgs doublet $H$ after electroweak symmetry breaking. For the phenomenological analysis, we denote by $\phi^{}_k$ the real scalar mode associated with the vacuum direction of the flavon multiplet $\Phi^{}_k$. Since our main purpose is to study the Higgs--flavon mixing and its implications for the leptogenesis parameter space, we consider a benchmark scalar sector characterized by the self-interaction of each flavon mode and its corresponding Higgs-portal interaction. The scalar potential is then written as
\begin{eqnarray}
V(H,\phi_k)
=
m_H^2H^\dagger H
+\lambda_H(H^\dagger H)^2
+\sum_{k}^{}
\left[
\frac{1}{2}\mu_k^2\phi_k^2
+\frac{1}{4}\lambda_k\phi_k^4
+\frac{1}{2}\lambda_{Hk}\phi_k^2H^\dagger H
\right].
\label{2.2.6}
\end{eqnarray}
Here, $\lambda^{}_k$ and $\lambda^{}_{Hk}$ denote the self-couplings of the effective flavon directions and the corresponding Higgs-portal couplings, respectively. The VEVs associated with these flavon directions are given by $u_1=v_1$, $u_2=\sqrt{2}v_2$, $u_3=\sqrt{3}v_3$, and $u_4=\sqrt{2}v_4$. The trilinear couplings relevant for leptogenesis arise from the Higgs-portal interactions and satisfy $\beta^{}_k=\lambda^{}_{Hk} u^{}_k$, where $\beta^{}_k$ denotes the dimensional trilinear coupling associated with the flavon direction $\phi_k$. For the numerical analysis, we consider a benchmark scenario with a degenerate flavon mass spectrum and universal trilinear couplings, i.e., $m_{\phi_k}=m_\phi$ and $\beta_k=\beta$. Under these assumptions, the scalar mixing associated with the different flavon directions can be parameterized by a common effective mixing angle $\theta$. The corresponding scalar couplings are then approximately given by
\begin{eqnarray}
&&\lambda_H
\simeq
\frac{m_h^2\cos^2\theta+m_\phi^2\sin^2\theta}{4v^2},
\nonumber\\
&&\lambda_k
\simeq
\frac{m_h^2\sin^2\theta+m_\phi^2\cos^2\theta}{2u_k^2},
\nonumber\\
&&\lambda_{Hk}
\simeq
\frac{(m_\phi^2-m_h^2)\sin2\theta}
{4\sqrt{2}\,v\,u_k}.
\label{2.2.7}
\end{eqnarray}

Theoretical consistency imposes several constraints on the scalar potential. Vacuum stability in each Higgs--flavon sector requires $\lambda^{}_H,\lambda^{}_k>0$ and $\lambda^{}_{Hk}>-2\sqrt{\lambda^{}_H\lambda^{}_k}$, ensuring that the potential remains bounded from below along the corresponding field directions. In addition, perturbativity requires all quartic couplings to remain within the perturbative regime. Although a naive upper limit is $4\pi$, we adopt the more conservative condition $|\lambda^{}_i|<3$, consistent with perturbative unitarity bounds~\cite{Goodsell:2018tti}. Beyond these theoretical requirements, the effective flavon sector is also constrained by Higgs measurements and collider searches. The observed signal strengths of the 125~GeV Higgs boson constrain the effective scalar mixing angle, leading to the upper bound $|\sin\theta|\lesssim0.3$~\cite{Robens:2015gla,Ilnicka:2018def}. Moreover, if $m^{}_{\phi}<m^{}_h/2$, the decay channels $h\to\phi^{}_k\phi^{}_k$ become kinematically accessible and contribute to the invisible decay width of the Higgs boson. The corresponding partial decay width for each flavon direction is given by
\begin{eqnarray}
\Gamma_{h\rightarrow\phi_k\phi_k}
=
\frac{\lambda_{Hk}^2v^2}{16\pi m_h}
\sqrt{1-\frac{4m_{\phi}^2}{m_h^2}},
\label{2.2.8}
\end{eqnarray}
where $m^{}_h=125$~GeV and the total SM Higgs width is $\Gamma^{\rm SM}_h\simeq4.07$~MeV~\cite{LHCHiggsCrossSectionWorkingGroup:2011wcg}. The corresponding invisible Higgs branching ratio is required to satisfy
$\mathrm{BR}(h\to\mathrm{inv})<0.107$ at the 95\% confidence level~\cite{ATLAS:2023tkt}.

\begin{figure*}
\centering
\includegraphics[width=6.5in]{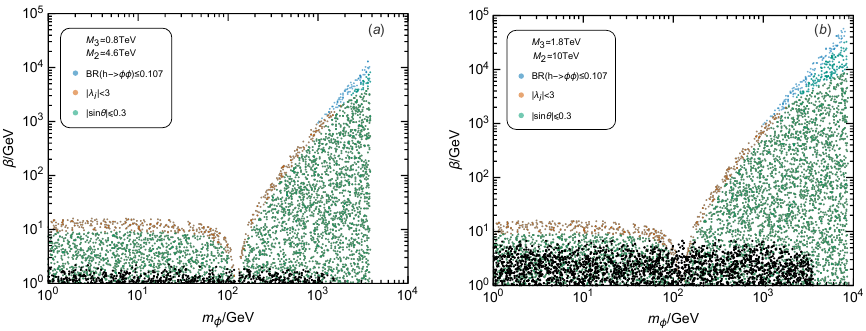}
\caption{
Viable parameter space of $\beta$ versus $m^{}_{\phi}$ for successful leptogenesis. The black points denote the parameter region that reproduces the observed baryon asymmetry, while the colored points indicate the regions satisfying the theoretical and experimental constraints.
}
\label{fig3}
\end{figure*}

Taking into account these constraints, in Figure~\ref{fig3} we show the parameter space of $m^{}_{\phi}$ versus $\beta$ that allows successful leptogenesis, together with the region allowed by the theoretical and experimental constraints discussed above. The viable parameter space is given by the overlap of these two regions. The results are obtained with the following parameter choices. For the neutrino mass squared differences and mixing angles, we adopt the global-fit results~\cite{global1,global2}, while the lightest neutrino mass is fixed at $m_1=0.001$~eV. The two Majorana CP phases of the neutrino mixing matrix are allowed to vary in the range $0$--$2\pi$, while the Dirac CP phase is constrained by the TM1 mixing pattern realized in this work. We take $|\alpha^{}_i|=10^{-1}$ as a representative benchmark, since larger values of $|\alpha^{}_i|$ generally lead to a broader viable parameter region. To obtain a representative and sufficiently comprehensive parameter space, we consider several benchmark values of $m^{}_{\rm D}$ within the region that successfully reproduces the observed baryon asymmetry of the Universe. The decay $N_2\rightarrow N_3\phi_{k,\rho}$ is required to be kinematically allowed, which imposes the condition $m^{}_\phi<M_2-M_3$. When this condition is not satisfied, the corresponding vertex and self-energy contributions involving $\phi^{}_{k,\rho}$ to the flavor-dependent CP asymmetries vanish, and the decay contribution $D_{23}$ is absent. The absence of viable points around $m^{}_{\phi}\sim m^{}_h$ originates from the near degeneracy between the Higgs boson and the flavon scalar states. In this region, it becomes difficult to simultaneously satisfy the reality and perturbativity requirements on the scalar potential parameters. We find that $m^{}_{\rm D}=0.2$~MeV, corresponding to $M_3\simeq0.8$~TeV and $M_2\simeq4.6$~TeV, provides a representative low-scale benchmark in which the observed $Y^{}_{\rm B}$ can be reproduced. For this choice, the viable parameter space extends up to $\beta\sim2$~GeV. As $m^{}_{\rm D}$ increases, larger values of $\beta$ become compatible with successful leptogenesis. For example, for $m^{}_{\rm D}=0.3$~MeV, corresponding to $M_3\simeq1.8$~TeV and $M_2\simeq10$~TeV, successful leptogenesis is realized for values of $\beta$ as large as $\sim6$~GeV.

\section{Leptogenesis for the case of $M^{}_{\rm R}\propto I$}

In this section, we study leptogenesis for the exactly degenerate form of $M^{}_{\rm R}$ given in Eq.~(\ref{1.3}), while the Dirac neutrino mass matrix $M^{}_{\rm D}$ acquires a nontrivial flavor structure through the spontaneous breaking of the flavor symmetry. For this form of $M^{}_{\rm R}$, the three RHNs are exactly degenerate in mass, causing the CP asymmetries of conventional RHN decays to vanish. We therefore investigate three-body RHN decays mediated by heavy vectorlike fermions, which provide a non-vanishing CP-asymmetry source even when the RHN masses are exactly degenerate. In Section~3.1, we introduce the corresponding framework and derive the CP asymmetries of these three-body RHN decays. In Section~3.2, we study the Boltzmann equations of the flavor-dependent lepton asymmetries and present the numerical results for the resulting baryon asymmetry.

\subsection{Model framework and CP asymmetries}

For nearly degenerate RHN masses, the CP asymmetries of RHN decays can be resonantly enhanced, and the contributions of the nearly degenerate RHNs to the final baryon asymmetry should be treated altogether~\cite{resonant1,resonant2}. However, for the case of $M^{}_{\rm R}\propto I$, the three RHNs are exactly degenerate in mass. In the conventional resonant leptogenesis mechanism, the CP asymmetries of RHN decays depend on the mass splittings and therefore vanish when $M^{}_I-M^{}_J=0$. Therefore, conventional resonant leptogenesis does not work for this form of $M^{}_{\rm R}$.

To realize flavon-driven leptogenesis in the exactly degenerate $M^{}_{\rm R}$ case, we introduce two heavy vectorlike fermion multiplets $F^{}_a$ ($a=1,2$), following the heavy fermion mediator mechanism employed for three-body RHN decays in Refs.~\cite{AristizabalSierra:2007ur}-\cite{AristizabalSierra:2009bh}. Through these heavy mediators, the flavon fields connect the RHN sector with the lepton--Higgs sector and allow the three-body RHN decays relevant for leptogenesis. Under the $S_4$ flavor symmetry considered here, the RHNs, lepton doublets, and vectorlike fermions transform as triplets of $S_4$, with $F^{}_a=(F^{}_{a1},F^{}_{a2},F^{}_{a3})$. In the chosen flavor basis, the couplings between the lepton doublets and the vectorlike fermions are diagonal, $(h_a)^{}_{\alpha\sigma}=h^{0}_a\delta^{}_{\alpha\sigma}$, where $\alpha$ and $\sigma$ denote the flavor indices of the lepton doublets and vectorlike fermions, respectively. Consequently, each lepton flavor $L^{}_\alpha$ couples only to the corresponding component $F^{}_{a\alpha}$ of the vectorlike-fermion triplet. The relevant interactions are given by
\begin{eqnarray}
-\mathcal{L}_{F} = \sum_{a}^{} \left[ \overline{F_a}\mathcal{M}_{F_a}F_a + \overline{L}\,h_a\,\widetilde H F_a + \sum_{k}^{} \overline{N}\, \lambda_a^{(k)} F_a\Phi_k \right] +\mathrm{h.c.},
\label{3.1.1}
\end{eqnarray}
where $\mathcal{M}^{}_{F_a}$ denotes the vectorlike-fermion mass matrix, while $h^{}_a$ and $\lambda^{(k)}_a$ denote the corresponding coupling matrices. In the flavor-symmetric limit, the vectorlike-fermion mass matrix takes the form $\mathcal{M}^{}_{F_a}=M^{}_{F_a}\mathbf{1}_{3\times3}$. For flavon multiplets with multiple components, the component index $\rho$ is retained in the effective couplings, following the notation and $S_4$ Clebsch--Gordan conventions introduced in Section~2.

Since the vectorlike fermions are much heavier than the RHNs, $M^{}_{F_a}\gg M_I$, they can be integrated out, giving rise to effective interactions among the RHNs, the lepton--Higgs sector, and the flavon fields. After the flavor symmetry is spontaneously broken, the flavon fields acquire the vacuum alignments given in Eq.~(\ref{2.1.2}), thereby generating the nontrivial flavor structure of the Dirac neutrino mass matrix:
\begin{eqnarray}
M_{\rm D}
=
v
\sum_{a,k,\rho}^{}
\frac{v_k}{M_{F_a}}
h_a
\lambda_a^{(k,\rho)\dagger}.
\label{3.1.2}
\end{eqnarray}
Since the coupling $h_a$ is flavor diagonal, the nontrivial flavor structure of $M_{\rm D}$ is generated by the couplings $\lambda_a^{(k,\rho)}$ together with the vacuum alignments of the flavon fields. In the RHN mass basis, the effective Yukawa coupling associated with the vectorlike fermion $F_a$ and the flavon component $\phi_{k,\rho}^{}$ is denoted by $Y_{\nu a}^{(k,\rho)}$. The total contribution from the two vectorlike fermion mediators is then written as $Y_\nu^{(k,\rho)} = Y_{\nu1}^{(k,\rho)}+Y_{\nu2}^{(k,\rho)}$.

\begin{figure*}
\centering
\includegraphics[width=5.5in]{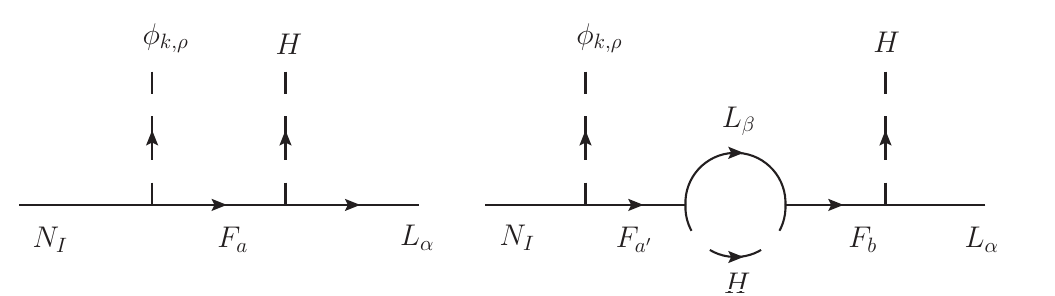}
\caption{Tree-level and self-energy diagrams contributing to the flavor-dependent CP asymmetry of $N_I\rightarrow L_\alpha H\phi_{k,\rho}$.}
\label{fig4}
\end{figure*}

In this scenario, we assume that leptogenesis takes place before the spontaneous breaking of the flavor symmetry. The flavon fields therefore have vanishing VEVs and enter the RHN decay processes as physical scalar states. Since the vectorlike fermions are much heavier than the RHNs, the two-body decay channels involving on-shell $F^{}_a$ are kinematically forbidden. However, the off-shell exchange of $F_a$ through the interactions in Eq.~(\ref{3.1.1}) induces the three-body decays $N_I\rightarrow L_\alpha H\phi_{k,\rho}$. As noted in Refs.~\cite{AristizabalSierra:2007ur,AristizabalSierra:2009tkm}, the vertex corrections vanish at the one-loop level, and the CP asymmetry arises solely from the interference between the tree-level and self-energy amplitudes shown in Figure~\ref{fig4}. Furthermore, for the flavor structure considered here, only the crossed self-energy contributions give rise to non-vanishing imaginary parts. The corresponding flavor-dependent CP asymmetry is then given by
\begin{eqnarray}
\varepsilon_{I\alpha}
=
\frac{3}{128\pi}
\frac{
\displaystyle
\sum_{k,\rho}
\frac{1}{v_k^2}
\left(
r_{1I}^{\,2}|h_1^0|^2-r_{2I}^{\,2}|h_2^0|^2
\right)
{\rm Im}
\left[
\left(Y_{\nu2}^{(k,\rho)}\right)_{\alpha I}^{*}
\left(Y_{\nu1}^{(k,\rho)}\right)_{\alpha I}
\right]
}
{
\displaystyle
\sum_{k,\rho}
\frac{1}{v_k^2}
\left(
Y_\nu^{(k,\rho)\dagger}
Y_\nu^{(k,\rho)}
\right)_{II}
}.
\label{3.1.3}
\end{eqnarray}
where $r^{}_{aI}=M^{}_I/M^{}_{F_a}$. The interference between the tree-level decay amplitude and the self-energy corrections involving different vectorlike fermion mediators generates a non-vanishing CP asymmetry. Unlike conventional resonant leptogenesis, this CP-violating source does not rely on the RHN mass splitting and therefore remains non-vanishing even in the exactly degenerate limit of $M^{}_{\rm R}$.

\subsection{Boltzmann equations and baryon asymmetry}

We now turn to the evolution of the generated lepton asymmetries. In the present study, the abundances of the three RHNs are evolved simultaneously in the Boltzmann equations. The CP asymmetries of the flavon-driven three-body RHN decays provide the source terms for the flavor-dependent lepton asymmetries. We extend the Boltzmann equations formulated in Ref.~\cite{AristizabalSierra:2009bh} to the present framework, and the corresponding evolution equations are given by
\begin{eqnarray}
&&\frac{dY_{N_I}}{dz}
=
-\frac{1}{sHz}
\left(
\frac{Y_{N_I}}{Y_{N_I}^{\rm eq}}-1
\right)
\gamma_{I}^{\rm tot},
\nonumber\\
&&\frac{dY_{\Delta L_{\alpha}}}{dz}
=
\sum_I
\frac{1}{sHz}
\left[
\varepsilon_{I\alpha}
\left(
\frac{Y_{N_I}}{Y_{N_I}^{\rm eq}}-1
\right)
\gamma_{I}^{\rm tot}
-
\frac{Y_{\Delta L_{\alpha}}}{Y_{\ell}^{\rm eq}}
\left\{
\gamma_{I\alpha}
+
\left(
\frac{Y_{N_I}}{Y_{N_I}^{\rm eq}}-1
\right)
\gamma(N_I\bar{L}_{\alpha}\rightarrow H\phi)
\right\}
\right].
\label{3.2.1}
\end{eqnarray}
Here, $Y^{}_{\Delta L_{\alpha}}=(n^{}_{L_\alpha}-n^{}_{\bar L_\alpha})/s$ denotes the flavor-dependent lepton asymmetry. The equilibrium abundances are given by $Y^{\rm eq}_{N_I}=\frac{45}{4\pi^4g_*}z^2K_2(z)$ and $Y^{\rm eq}_{\ell}=\frac{135\zeta(3)}{4\pi^4g_*}$. The first term in the equation for $Y_{\Delta L_\alpha}$ represents the source term for the flavor-dependent lepton asymmetries generated by the CP asymmetries of the flavon-driven three-body RHN decays, while the second term describes the washout effects from inverse decays and $\Delta L=1$ scattering processes. The reaction densities appearing in the Boltzmann equations are defined as
\begin{eqnarray}
&& \gamma_{I\alpha}
=
\sum_{k,\rho}
\left[
\gamma(N_I\to L_\alpha H\phi_{k,\rho})
+\gamma(N_I\bar{\phi}_{k,\rho}\to L_\alpha H)
+\gamma(N_I\bar{H}\to L_\alpha\phi_{k,\rho})
+\gamma(N_I\bar{L}_\alpha\to H\phi_{k,\rho})
\right],
\nonumber\\
&& \gamma_{I}^{\rm tot}
=
\sum_{\alpha}
\left(
\gamma_{I\alpha}
+
\bar{\gamma}_{I\alpha}
\right),
\label{3.2.2}
\end{eqnarray}
where $\gamma^{}_{I\alpha}$ denotes the total reaction density of the decay and scattering processes involving lepton flavor $\alpha$, while $\gamma^{\rm tot}_I$ denotes the total reaction density summed over lepton flavors and the corresponding CP-conjugate processes. The explicit expressions for these reaction densities are given in Appendix~A.2.

As is known, depending on the temperature range in which leptogenesis takes place (approximately the RHN mass scale), the following three distinct leptogenesis regimes can be identified~\cite{flavor1, flavor2}. For $M_0\gtrsim10^{12}\ {\rm GeV}$, the charged-lepton Yukawa interactions are not in thermal equilibrium, and the lepton flavors are therefore indistinguishable. Consequently, the three lepton flavors should be treated universally. For $10^9\ {\rm GeV}\lesssim M_0\lesssim10^{12}\ {\rm GeV}$, the $y^{}_\tau$-mediated interactions are in thermal equilibrium, whereas those associated with $e$ and $\mu$ are not. Therefore, the $\tau$ flavor is distinguishable from the other two, resulting effectively in two flavor sectors: the $\tau$ flavor and a coherent superposition of the $e$ and $\mu$ flavors. For $M_0\lesssim10^9\ {\rm GeV}$, the $y^{}_\mu$-mediated interactions also enter thermal equilibrium, so that all three flavors become distinguishable and must be treated separately. In the numerical analysis, the corresponding flavor regime is selected according to the value of $M_0$. The final baryon asymmetry is obtained as $Y^{}_{\rm B}=c\sum_\alpha Y^{}_{\Delta L_\alpha}$.

\begin{figure*}
\centering
\includegraphics[width=6.5in]{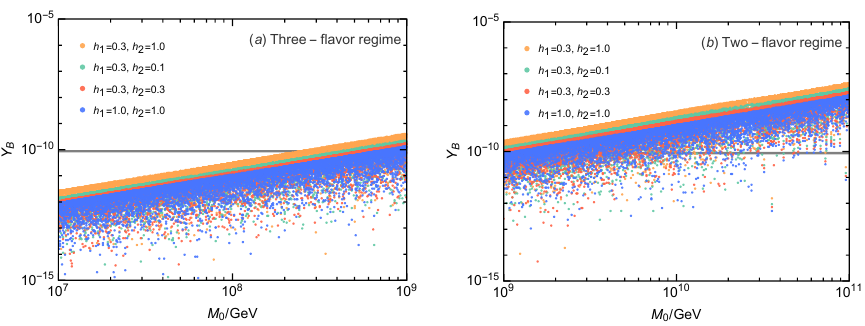}
\caption{For the three- and two-flavor regimes, the allowed values of $Y^{}_{\rm B}$ as functions of $M_0$ for some benchmark values of $h^{}_1$ and $h^{}_2$. The horizontal line stands for the observed value of $Y^{}_{\rm B}$. }
\label{fig5}
\end{figure*}

The three-body decay channels $N_I\rightarrow L^{}_\alpha H\phi_{k,\rho}$ provide a non-vanishing CP-asymmetry source in the exactly degenerate form of $M^{}_{\rm R}$, as shown in Eq.~(\ref{3.1.3}). We therefore investigate whether this CP-violating source can lead to successful leptogenesis. Owing to the special form of $M^{}_{\rm R}$ in Eq.~(\ref{1.3}), taking $M_0$ as an input parameter allows us to determine the Dirac neutrino mass matrix $M^{}_{\rm D}$ from the experimental results for the neutrino parameters. Since $M^{}_{\rm D}$ receives contributions from both $F_1$ and $F_2$, their relative contributions are parameterized by a phase $\theta^{}_{\rm F}$. In Figure~\ref{fig5}(a) and (b) (for the three- and two-flavor regimes, respectively), we show the resulting baryon asymmetry $Y^{}_{\rm B}$ as a function of $M_0$ for several benchmark choices of $h_1$ and $h_2$ (see the figure for details). The results are obtained with the following parameter choices. For the neutrino mass squared differences and mixing angles, we adopt the global-fit results in Refs.~\cite{global1,global2}. The two Majorana CP phases and $\theta_{\rm F}$ are allowed to vary in the range $0$--$2\pi$, while the Dirac CP phase is constrained by the TM1 mixing pattern realized in this model. As in the previous section, we consider the normal ordering of light-neutrino masses, with the lightest neutrino mass varied in the range $0.001~{\rm eV}\leq m_1\leq0.1~{\rm eV}$. The couplings associated with the $NF\Phi_k$ interactions are taken to be $0.1$ as representative benchmark values, while the heavy fermion masses are chosen as $M_{F_2}=2M_{F_1}=10M_0$.

The results show that the observed value of $Y^{}_{\rm B}$ can be successfully reproduced in both flavor regimes. It is clear that the predicted baryon asymmetry increases with $M_0$, while different choices of $h_1$ and $h_2$ mainly affect the magnitude of $Y^{}_{\rm B}$. For fixed $h_1=0.3$, different values of $h_2$ lead to different values of $Y^{}_{\rm B}$. This behavior can be understood from the factor $r_1^2|h_1|^2-r_2^2|h_2|^2$ appearing in the CP asymmetry. Since $M_{F_2}=2M_{F_1}$, the two contributions approximately cancel around $h_2=0.6$. Therefore, increasing $h_2$ from $0.1$ to $0.3$ reduces $Y^{}_{\rm B}$ due to the increasing cancellation between the two contributions, while for $h_2=1.0$ the magnitude of the CP asymmetry increases again beyond the cancellation region. For $h_1=h_2$, increasing the common coupling from $0.3$ to $1.0$ leads to a smaller final $Y^{}_{\rm B}$ due to the enhanced washout effects. Overall, the observed baryon asymmetry can be reproduced for $M_0$ as low as $2\times10^{8}\ {\rm GeV}$.

\section{Summary}

While the type-I seesaw model provides an attractive framework for simultaneously explaining the origin of neutrino masses and the baryon asymmetry of the Universe through leptogenesis, flavor symmetries offer a natural approach to understanding the observed neutrino mixing pattern. In many type-I seesaw models based on non-Abelian flavor symmetries, the three lepton doublets and the three right-handed neutrinos are assigned to triplet representations. In the exact flavor-symmetric limit, this assignment can lead to highly constrained neutrino mass matrices, including the exactly degenerate forms of $M^{}_{\rm D}$ or $M^{}_{\rm R}$ shown in Eq.~(\ref{1.3}). However, these exactly degenerate structures prevent the conventional leptogenesis mechanism from working, either due to the orthogonality relations among different columns of the neutrino Yukawa coupling matrix or because of the exact mass degeneracy of the three RHNs.

In this paper, we have investigated how leptogenesis can be realized for the exactly degenerate forms of $M_{\rm D}$ and $M_{\rm R}$ while preserving the flavor structures imposed by the underlying flavor symmetry. We have shown that the flavon fields responsible for generating the nontrivial neutrino flavor structure can also participate in RHN dynamics and provide the CP-violating sources required for successful leptogenesis. In this way, leptogenesis naturally emerges from the same flavon sector responsible for the neutrino flavor structure, establishing a direct connection between neutrino flavor symmetry and the origin of the baryon asymmetry. To demonstrate this mechanism, we have considered an $S_4$ flavor framework realizing the phenomenologically viable TM1 mixing pattern.

For the case of $M^{}_{\rm D}\propto I$, the flavon--RHN interactions open additional RHN decay channels $N_I\rightarrow N_J \phi_{}$ and generate new CP-violating contributions to RHN decays. The observed baryon asymmetry can be reproduced in a parameter region consistent with relevant theoretical and experimental constraints. In particular, for $m^{}_{\rm D}=0.2$~MeV, the lightest RHN mass can be as low as $M_3\simeq0.8$~TeV, with the viable parameter space extending to $\beta\sim2$~GeV. Larger values of $m^{}_{\rm D}$ allow correspondingly larger values of $\beta$.

For the case of $M^{}_{\rm R}\propto I$, three-body decays $N_I\rightarrow L_\alpha H\phi_{}$ mediated by heavy vectorlike fermions provide a CP asymmetry source even when the RHN masses are exactly degenerate. The observed baryon asymmetry can be reproduced in both the two- and three-flavor regimes. The resulting $Y^{}_{\rm B}$ generally increases with $M_0$, while its magnitude depends on the choices of $h_1$ and $h_2$ through their contributions to the CP asymmetry and washout effects. Successful leptogenesis can be realized for $M_0$ as low as about $2\times10^{8}$~GeV.

In summary, our results demonstrate that the flavon fields can provide a viable way to overcome the obstruction to conventional leptogenesis caused by the exactly degenerate flavor structures considered here. Although our numerical analysis has been performed within an $S_4$ realization of TM1 mixing, the underlying mechanism is not restricted to this particular mixing pattern and may be applied more generally to flavor-symmetry type-I seesaw models in which flavon fields participate in RHN dynamics.

\appendix

\section{Scattering functions and reaction densities}

\subsection{Flavon-mediated $\Delta L=2$ scattering}

For completeness, we present the flavon-mediated $\Delta L=2$ scattering processes included in the Boltzmann equations, which contribute to the washout of the RHN abundances. The corresponding scattering function for the process $N_I N_J\rightarrow HH$ is given by
\begin{eqnarray}
S_{N_I N_J\rightarrow HH}(z)
=
\frac{M_I}{64\pi^2 H(T=M_I)}
\int_{w_{\rm min}}^{\infty}
{\rm d}w\,
\sqrt{w}\,
{\cal K}_1(\sqrt{w})\,
\widehat{\sigma}_{N_I N_J\rightarrow HH}
\left(
\frac{wM_I^2}{z_I^2}
\right),
\label{A.1.1}
\end{eqnarray}
where $w=s/T^2$, $z_I=M_I/T$, and $w_{\rm min}=(M_I+M_J)^2/T^2$. Including the contributions of all relevant flavon mediators, the reduced cross section can be expressed as~\cite{Alanne:2017sip}
\begin{eqnarray}
\widehat{\sigma}_{N_I N_J\rightarrow HH}(s)
=
\frac{\beta^2}{32\pi}
\sqrt{\left(s-M_I^2-M_J^2\right)^2-4M_I^2M_J^2}\,
\frac{s-(M_I+M_J)^2}{s}
\left|
\sum_{k,\rho}
\frac{\alpha^{({k,\rho})}_{IJ}}
{s-m_{\phi_k}^2+i m_{\phi_k}\Gamma_{\phi_{k,\rho}}}
\right|^2 .
\label{A.1.2}
\end{eqnarray}
where $\Gamma^{}_{\phi_{k,\rho}}$ denotes the total decay width of the flavon field $\phi^{}_{k,\rho}$. In the parameter region considered here, it receives contributions from the decay into Higgs doublets and, when kinematically allowed, the decay into an RHN pair $\phi^{}_{k,\rho}\rightarrow N_3N_3$,
\begin{eqnarray}
\Gamma_{\phi_{k,\rho}}
=
\Gamma(\phi_{k,\rho}\rightarrow HH)
+
\Gamma(\phi_{k,\rho}\rightarrow N_3N_3),
\label{A.1.3}
\end{eqnarray}
where the second term is included only when $m^{}_{\phi_k}>2M_3$. The decays of $\phi^{}_{k,\rho}$ into heavier RHN pairs are kinematically forbidden by the RHN mass hierarchy considered here. The contributions of different flavon mediators are summed coherently at the amplitude level, so that their interference effects are automatically included in the reduced cross section in Eq.~(\ref{A.1.2}).

\subsection{Decay and $\Delta L=1$ scatterings}

The reaction densities entering the Boltzmann equations in Section~3.2 arise from the flavon-driven three-body RHN decays and the associated $\Delta L=1$ scattering processes mediated by the heavy vectorlike fermions. For each three-body decay channel, the corresponding partial decay width is given by~\cite{AristizabalSierra:2009bh}
\begin{eqnarray}
\Gamma(N_I\rightarrow L_\alpha H\phi_{k,\rho})
=
\frac{M_I^3}{192\pi^3 v_k^2}
\left|
\left(Y_{\nu}^{(k,\rho)}\right)_{\alpha I}
\right|^2 .
\label{A.2.1}
\end{eqnarray}
The corresponding thermally averaged decay reaction density is then given by
\begin{eqnarray}
\gamma(N_I\rightarrow L_\alpha H\phi_{k,\rho})
=
n_{N_I}^{\rm eq}
\frac{K_1(z)}{K_2(z)}
\Gamma(N_I\rightarrow L_\alpha H\phi_{k,\rho}).
\label{A.2.2}
\end{eqnarray}

The scattering contributions to the washout terms arise from the processes $N_I\bar{\phi}_{k,\rho}\rightarrow L_\alpha H$, $N_I\bar H\rightarrow L_\alpha\phi_{k,\rho}$, and $N_I\bar L_\alpha\rightarrow H\phi_{k,\rho}$. For a generic $2\rightarrow2$ scattering process, the thermally averaged reaction density is given by
\begin{eqnarray}
\gamma(a b\rightarrow c d)
=
\frac{M_{I}^{4}}{512\pi^{5}z}
\int_{1}^{\infty}
dx\,
\sqrt{x}\,
K_1\!\left(z\sqrt{x}\right)
\widehat{\sigma}(x),
\label{A.2.3}
\end{eqnarray}
where $\widehat{\sigma}(x)$ denotes the reduced cross section. In the present framework, it can be written as
\begin{eqnarray}
&& \hat{\sigma}^{({k,\rho})}(x)=
r_1^{2}
\left|
h_1^0
(\lambda_1^{({k,\rho})})_{I\alpha}
\right|^2
F_1(x)
+
r_2^{2}
\left|
h_2^0
(\lambda_2^{({k,\rho})})_{I\alpha}
\right|^2
F_2(x)
\nonumber\\
&&\hspace{1.7cm}
+
2r_1r_2
\operatorname{Re}
\left[
(h_1^0)^{*}h_2^0
(\lambda_1^{({k,\rho})})_{I\alpha}^{*}
(\lambda_2^{({k,\rho})})_{I\alpha}
\right]
G_{1,2}(x).
\label{A.2.4}
\end{eqnarray}
Since the vectorlike fermions transform as flavor triplets, the diagonal form $(h_a)^{}_{\alpha\sigma}=h^0_a\delta^{}_{\alpha\sigma}$ implies that a lepton of flavor $\alpha$ couples only to the corresponding component $F^{}_{a\alpha}$. The interaction associated with this flavor channel is therefore governed by the coupling $(\lambda_a^{({k,\rho})})_{I\alpha}$. The functions $F^{}_a(x)$ and $G^{}_{a,b}(x)$ describe the contributions of the individual vectorlike fermions and their interference, respectively, including the corresponding $s$-, $t$-, and $u$-channel kinematics. In the limit $M^{}_{F_a}\gg M_I$, the $t$- and $u$-channel processes can be described by effective point-like interactions obtained after integrating out the heavy vectorlike fermions. The corresponding reaction densities for $N_I\bar H\rightarrow L_\alpha\phi_{k,\rho}$ and $N_I\bar L_\alpha\rightarrow H\phi_{k,\rho}$ are given by
\begin{eqnarray}
&& \gamma^{({k,\rho})}_{t,u}
=
\frac{M_I^4}{1024\pi^5 z}
\left(
r_1^{2}
\left|
h_1^0
(\lambda_1^{({k,\rho})})_{I\alpha}
\right|^2
+
r_2^{2}
\left|
h_2^0
(\lambda_2^{({k,\rho})})_{I\alpha}
\right|^2
+
2r_1r_2
\operatorname{Re}
\left[
(h_1^0)^*h_2^0
(\lambda_1^{({k,\rho})})_{I\alpha}^{*}
(\lambda_2^{({k,\rho})})_{I\alpha}
\right]
\right)
f(z),
\nonumber\\
\label{A.2.5}
\end{eqnarray}
with
\begin{eqnarray}
f(z)
=
\int_{1}^{\infty}
dx\,
\frac{x^{2}-1}{\sqrt{x}}
K_1\left(z\sqrt{x}\right).
\label{A.2.6}
\end{eqnarray}
For the $s$-channel scattering process $N_I\bar{\phi}^{}_{k,\rho}\rightarrow L_\alpha H$, the point-like approximation is not sufficiently accurate, and the full kinematic dependence is therefore retained. The corresponding functions $F_a(x)$ and $G_{a,b}(x)$ are given by
\begin{eqnarray}
G_{s}^{a,b}(x)
=
\frac{x-1}{H_{s,1}^{a,b}(x)}
\left[
(1-x)H_{s,2}^{a,b}(x)
+
(1+x)H_{s,3}^{a,b}(x)
\right],
\label{A.2.7}
\end{eqnarray}
with
\begin{eqnarray}
&& H_{s,1}^{a,b}(x)
=
2x
\left(
1-r_a^{2}x-2ir_a\eta_{a\alpha}
\right)
\left(
1-r_b^{2}x+2ir_b\eta_{b\alpha}
\right),
\nonumber \\
&& H_{s,2}^{a,b}(x)
=
2r_a r_b+r_a+r_b
+
\mathrm{i}r_a r_b
\left(
\eta_{a\alpha}-\eta_{b \alpha}
\right)
+
r_ar_b(x-1),
\nonumber \\
&& H_{s,3}^{a,b}(x)
=
r_ar_b+r_a+r_b
+ir_ar_b(\eta_{a\alpha}-\eta_{b\alpha})
+
(1+ir_a\eta_{a\alpha})
(1-ir_b\eta_{b\alpha}),
\label{A.2.8}
\end{eqnarray}
and
\begin{eqnarray}
F_{s}^{a}(x)
=
G_{s}^{a,a}(x)
\label{A.2.9}
\end{eqnarray}
The finite-width effects of the intermediate vectorlike fermions are parameterized by the dimensionless quantity $\eta^{}_{a\alpha}\equiv\Gamma^{}_{F_{a\alpha}}/M_0$, where $\Gamma^{}_{F_{a\alpha}}$ denotes the total decay width of $F^{}_{a\alpha}$. Its explicit expression is
\begin{eqnarray}
\eta_{a\alpha}
=
\frac{1}{8\pi r_a}
\left[
(1-r_a^2)(1+r_a)^2
\sum_{I,k,\rho}
\left|
(\lambda_a^{(k,\rho)})_{I\alpha}
\right|^2
+
\frac12 |h_a^0|^2
\right].
\label{A.2.10}
\end{eqnarray}

\vspace{0.5cm}

\underline{Acknowledgments} \vspace{0.2cm}

This work was supported in part by the National Natural Science Foundation of China under Grant Nos.~12475112 and 12447101, Liaoning Revitalization Talents Program under Grant No.~XLYC2403152, and the Basic Research Business Fees for Universities in Liaoning Province under Grant No.~LJ212410165050.


\begin{thebibliography}{99}

\bibitem{xing} Z. Z. Xing, Phys. Rep. {\bf 854}, 1 (2020).

\bibitem{planck} P. A. R. Ade {\it et al.} (Planck Collaboration), Astron. Astrophys. A {\bf16}, 571 (2014).

\bibitem{seesaw1} P. Minkowski, Phys. Lett. B {\bf 67}, 421 (1977).

\bibitem{seesaw2} M. Gell-Mann, P. Ramond and R. Slansky, in Supergravity, edited by P. van Nieuwenhuizen and D. Freedman, (North-Holland, 1979), p. 315.

\bibitem{seesaw3}  T. Yanagida, in Proceedings of the Workshop on the Unified Theory and the Baryon Number in the Universe, edited by O. Sawada and A. Sugamoto (KEK Report No. 79-18, Tsukuba, 1979), p. 95.

\bibitem{seesaw4} R. N. Mohapatra and G. Senjanovic, Phys. Rev. Lett. {\bf 44}, 912 (1980).

\bibitem{seesaw5} J. Schechter and J. W. F. Valle, Phys. Rev. D {\bf22}, 2227 (1980).

\bibitem{leptogenesis} M. Fukugita and T. Yanagida, Phys. Lett. B {\bf 174}, 45 (1986).

\bibitem{Lreview1} W. Buchmuller, R. D. Peccei and T. Yanagida, Ann. Rev. Nucl. Part. Sci. {\bf 55}, 311 (2005).

\bibitem{Lreview2} W. Buchmuller, P. Di Bari and M. Plumacher, Annals Phys. {\bf 315}, 305 (2005).

\bibitem{Lreview3} S. Davidson, E. Nardi and Y. Nir, Phys. Rept. {\bf 466}, 105 (2008).

\bibitem{Lreview4} D. Bodeker and W. Buchmuller, Rev. Mod. Phys. {\bf 93}, 035004 (2021).

\bibitem{FS1} G. Altarelli and F. Feruglio, Rev. Mod. Phys. {\bf 82}, 2701 (2010).

\bibitem{FS2} S. F. King and C. Luhn, Rept. Prog. Phys. {\bf 76}, 056201 (2013).

\bibitem{FS3} F. Feruglio and A. Romanino, Rev. Mod. Phys. {\bf 93}, 015007 (2021).

\bibitem{FS4} G. J. Ding and S. F. King, Rept. Prog. Phys. {\bf 87} 084201 (2024).

\bibitem{King:2011zj} S.~F.~King and C.~Luhn, JHEP \textbf{09}, 042 (2011).

\bibitem{Luhn:2013lkn} C.~Luhn, Nucl. Phys. B \textbf{875}, 80 (2013).

\bibitem{Ding:2013hpa} G.~J.~Ding, S.~F.~King, C.~Luhn and A.~J.~Stuart, JHEP \textbf{05}, 084 (2013).

\bibitem{TB2-TM} Z. Z. Xing, Phys. Lett. B {\bf 533}, 85 (2002).

\bibitem{TM-1} J. D. Bjorken, P. F. Harrison and W. G. Scott, Phys. Rev. D {\bf 74}, 073012 (2006).

\bibitem{TM-2} Z. Z. Xing and S. Zhou, Phys. Lett. B {\bf 653}, 278 (2007).

\bibitem{TM-3} X. G. He and A. Zee, Phys. Lett. B {\bf 645}, 427 (2007).

\bibitem{TM-4} C. H. Albright and W. Rodejohann, Eur. Phys. J. C {\bf 62}, 599 (2009).

\bibitem{TM-5} C. H. Albright, A. Dueck and W. Rodejohann, Eur. Phys. J. C {\bf 70}, 1099 (2010).

\bibitem{TB1} P. F. Harrison, D. H. Perkins and W. G. Scott, Phys. Lett. B {\bf 530}, 167 (2002).

\bibitem{global1} I.~Esteban, M.~C.~Gonzalez-Garcia, M.~Maltoni, I.~Martinez-Soler, J.~P.~Pinheiro and T.~Schwetz, JHEP \textbf{12}, 216 (2024).

\bibitem{global2} F.~Capozzi, W.~Giar{\`e}, E.~Lisi, A.~Marrone, A.~Melchiorri and A.~Palazzo, Phys. Rev. D \textbf{111}, 093006 (2025).

\bibitem{JUNO:2025gmd} A.~Abusleme \textit{et al.} [JUNO], Nature \textbf{654} no.8118, 343 (2026).

\bibitem{Xing:2025bdm} Z.~z.~Xing, Sci. Bull. \textbf{71}, no.8, 1899 (2026).

\bibitem{He:2025idv} X.~G.~He, Phys. Lett. B \textbf{874}, 140270 (2026).

\bibitem{Zhang:2025jnn} D.~Zhang, Phys. Rev. D \textbf{113}, no.5, 055035 (2026).

\bibitem{Jiang:2025hvq} W.~H.~Jiang, R.~Ouyang and Y.~L.~Zhou, Phys. Rev. D \textbf{114} no.5, 055031 (2026).

\bibitem{Capozzi:2025ovi} F.~Capozzi, E.~Lisi, F.~Marcone, A.~Marrone and A.~Palazzo, Phys. Rev. D \textbf{114} no.1, 016026 (2026).

\bibitem{Ding:2025dzc} G.~J.~Ding, R.~Kumar, N.~Nath, R.~Srivastava and J.~W.~F.~Valle, Eur. Phys. J. C \textbf{86} no.9, 1037 (2026).

\bibitem{Ardakanian:2026rwz} N.~Ardakanian, [arXiv:2603.21264 [hep-ph]].

\bibitem{LeDall:2014too} M.~Le Dall and A.~Ritz, Phys. Rev. D \textbf{90}, no.9, 096002 (2014).

\bibitem{Alanne:2017sip} T.~Alanne, A.~Meroni and K.~Tuominen, Phys. Rev. D \textbf{96}, no.9, 095015 (2017).

\bibitem{Alanne:2018brf} T.~Alanne, T.~Hugle, M.~Platscher and K.~Schmitz, JCAP \textbf{03}, 037 (2019).

\bibitem{Barreiros:2022fpi} D.~M.~Barreiros, H.~B.~C{\^a}mara, R.~G.~Felipe and F.~R.~Joaquim, JHEP \textbf{01}, 010 (2023).

\bibitem{Abe:2021mfy} Y.~Abe, T.~Ito and K.~Yoshioka, JHEP \textbf{01}, 019 (2023).

\bibitem{Ahmed:2025gww} A.~Ahmed, J.~P.~Garc{\'e}s and M.~Lindner, Phys. Rev. D \textbf{112}, no.3, 035026 (2025).

\bibitem{Ghosh:2024mpz} D.~K.~Ghosh, P.~Ghosh, K.~Mukherjee and N.~Narendra, Eur. Phys. J. C \textbf{85}, no.10, 1217 (2025).

\bibitem{Spalding:2026otj} A.~Spalding, [arXiv:2609.24990 [hep-ph]].

\bibitem{Shao:2026xcb} Y.~Shao and Z.~h.~Zhao, [arXiv:2604.20581 [hep-ph]].

\bibitem{Goodsell:2018tti} M.~D.~Goodsell and F.~Staub, Eur. Phys. J. C \textbf{78}, no.8, 649 (2018).

\bibitem{Robens:2015gla} T.~Robens and T.~Stefaniak, Eur. Phys. J. C \textbf{75}, 104 (2015).

\bibitem{Ilnicka:2018def} A.~Ilnicka, T.~Robens and T.~Stefaniak, Mod. Phys. Lett. A \textbf{33}, no.10n11, 1830007 (2018).

\bibitem{LHCHiggsCrossSectionWorkingGroup:2011wcg} S.~Dittmaier \textit{et al.} [LHC Higgs Cross Section Working Group], [arXiv:1101.0593 [hep-ph]].

\bibitem{ATLAS:2023tkt} G.~Aad \textit{et al.} [ATLAS], Phys. Lett. B \textbf{842}, 137963 (2023).

\bibitem{resonant1} A. Pilaftsis, Phys. Rev. D {\bf 56}, 5431 (1997).

\bibitem{resonant2} A. Pilaftsis and T. E. J. Underwood, Nucl. Phys. B {\bf 692}, 303 (2004).

\bibitem{AristizabalSierra:2007ur} D.~Aristizabal Sierra, M.~Losada and E.~Nardi, Phys. Lett. B \textbf{659}, 328 (2008).

\bibitem{AristizabalSierra:2009tkm} D.~Aristizabal Sierra, L.~A.~Munoz and E.~Nardi, J. Phys. Conf. Ser. \textbf{171}, 012078 (2009).

\bibitem{AristizabalSierra:2009bh} D.~Aristizabal Sierra, L.~A.~Munoz and E.~Nardi, Phys. Rev. D \textbf{80}, 016007 (2009).

\bibitem{flavor1} A. Abada, S. Davidson, F. X. Josse-Michaux, M. Losada and A. Riotto, JCAP {\bf0604}, 004 (2006).

\bibitem{flavor2} E. Nardi, Y. Nir, E. Roulet and J. Racker, JHEP {\bf0601}, 164 (2006).





\end{thebibliography}
\end{document}